\documentclass[aps,prb,twocolumn,superscriptaddress,showpacs,longbibliography]{revtex4-1}

\usepackage{amsmath,amssymb,graphicx,subfigure}

\usepackage[utf8]{inputenc}
\usepackage[T1]{fontenc}
\usepackage{xcolor}
\usepackage[normalem]{ulem}
\usepackage{multirow}
\usepackage{stmaryrd}
\usepackage{dblfloatfix} 

\IfFileExists{newtxtext.sty}
   {\usepackage{newtxtext,newtxmath}}
   {\IfFileExists{stix.sty}
      {\usepackage{stix}}
      {\IfFileExists{mathptmx.sty}
      {\usepackage{mathptmx}}{} } }

\usepackage{textcomp}

\usepackage{bm}

\IfFileExists{siunitx.sty}{\usepackage{booktabs,siunitx}}{}

\pdfoutput=1
\usepackage{color}
\definecolor{LinkColor}{rgb}{0.256,0.439,0.588}
\usepackage{hyperref}
\hypersetup{
   colorlinks=true,
   citecolor=LinkColor,
   linkcolor=LinkColor,
   urlcolor=LinkColor
}

\newcommand{\bra}[1]{\langle#1\rvert}
\newcommand{\ket}[1]{\lvert#1\rangle}

\newcommand{\be}{\begin{equation}}
\newcommand{\ee}{\end{equation}}
\newcommand{\bea}{\begin{eqnarray}}
\newcommand{\eea}{\end{eqnarray}}

\usepackage{pifont}

\begin{document}

\title{Green's Function Approach to Interacting Higher-order Topological Insulators}

\author{Heqiu Li}
	\affiliation{Department of Physics, University of Toronto, Toronto, Ontario M5S 1A7, Canada}

\author{Hae-Young Kee}
\email{hykee@physics.utoronto.ca}
	\affiliation{Department of Physics, University of Toronto, Toronto, Ontario M5S 1A7, Canada}
	\affiliation{Canadian Institute for Advanced Research, CIFAR Program in Quantum Materials, Toronto, Ontario M5G 1M1, Canada}

\author{Yong Baek Kim}
\email{ybkim@physics.utoronto.ca}
	\affiliation{Department of Physics, University of Toronto, Toronto, Ontario M5S 1A7, Canada}
	\affiliation{School of Physics, Korea Institute for Advanced Study, Seoul 02455, Korea}

\date{\today}

\begin{abstract}

The Bloch wave functions have been playing a crucial role in the diagnosis of topological phases in non-interacting systems. However, the Bloch waves are no longer applicable in the presence of finite Coulomb interaction and alternative approaches are needed to identify the topological indices. In this paper, we focus on three-dimensional higher-order topological insulators protected by $C_4T$ symmetry and show that the topological index can be computed through eigenstates of inverse Green's function at zero frequency. If there is an additional $S_4$ rotoinversion symmetry, the topological index $P_3$ can be determined by eigenvalues of $S_4$ at high symmetry momenta, similar to the Fu-Kane parity criterion. We verify this method using many-body exact diagonalization in higher-order topological insulators with interaction. We also discuss the realization of this higher-order topological phase in tetragonal lattice structure with $C_4T$-preserving magnetic order. Finally, we discuss the boundary conditions necessary for the hinge states to emerge and show that these hinge states exist even when the boundary is smooth and without a sharp hinge. 

\end{abstract}

\maketitle

\section{Introduction}

Topological phases of matter are characterized by an exotic bulk-boundary correspondence that enforces the boundary to be gapless although the bulk has a finite energy gap. The first-order topological insulators in $d$-dimension have $(d-1)$-dimensional in-gap boundary states that are robust to perturbations as long as the bulk gap remains open and the symmetries are not broken by perturbations~\cite{Qi_TI,Qi2008,Hasan_TI,FuKane_Z2,FuKaneMele3D,FuKane_inv,KaneMele2005}. The higher-order topological insulators have in-gap states at $(d-n)$-dimensional boundary~\cite{Benalcazar2017,Benalcazar2017b,Bernevig_HOTI,Ahn2019,Wieder2018,Ezawa2018,Ezawa2018b,Fang2015,Van2018,Khalaf2018,Kooi2018,Calugaru2019,Varjas2015,Ezawa2019,Wang2018,Song2017,Matsugatani2018,Langbehn2017,Yue2019,Hsu2018,Queiroz2018,Xue2019,Geier2018,Schindler2018,Trifunovic2019,Ghorashi2019,Nag2021,Ghosh2021,Trifunovic2021,Benalcazar2019,Fang2021,Lee2022}, e.g., the second-order topological insulators in three-dimensional space have gapless hinge states. These nontrivial topological features are indicated by topological indices. For non-interacting systems, the Bloch wave functions have been playing an important role in the diagnosis of topological phases. From the Berry curvature for Chern insulators to the nested Wilson loop~\cite{Bernevig_HOTI,Benalcazar2017,Benalcazar2017b,Yu2011,Franca2018,Adrien2018} for higher order topological insulators, all these quantities involve Bloch wave functions. The representation of Bloch wave functions under symmetry groups also enables a highly efficient approach to identify topological phases regardless of the microscopic details in materials~\cite{Fang2012,topchem,indicator,Kruthoff2017,S4index,Ono2018,Song2018b,Zhang2019,Tang2019,Vergniory2019}.

The presence of Coulomb interaction poses several challenges in characterizing the topological properties of electronic systems. Firstly, some of the topological phases may not be stable under interaction, and interaction can modify the topological classification. For example, the topological classification of one-dimensional Majorana chain can be reduced from $\mathbb Z$ to $\mathbb Z_8$ by interaction~\cite{KitaevZ8}. Secondly, topological indices in non-interacting electronic systems are usually defined in terms of Bloch wave functions, which can only capture information in the single-particle Hamiltonian and cannot describe correlation effects. Therefore, in the presence of interaction it is desirable to find an alternative approach that can take into account the many-body physics and characterize the topological properties of the interacting system~\cite{Resta1998,Slager2015,Shiozaki2018,Kang2019,Wheeler2019,Kudo2019,Kang2021}.


The well-known Chern insulators and time-reversal protected $\mathbb Z_2$ topological insulators are examples of first-order topological insulators that are known to be robust under weak Coulomb interaction which does not close the band gap~\cite{Qi2008}. Furthermore, Wang, Qi and Zhang~\cite{Wang2012inv,Wang2012int,Wang2012GSC} suggested when there is finite interaction in these systems, the role of Bloch wave functions can be played by the eigenstates of the inverse Green's function at zero frequency such that the topological indices can be formulated through these eigenstates. 

For higher-order topological phases, however, the fate of topological features under interaction becomes less clear. For example, Ref. \onlinecite{HOTIC1} questioned the stability of the three-dimensional $C_4T$-protected second order topological insulator~\cite{Bernevig_HOTI} under weak Coulomb interaction {($C_4$ is fourfold rotation and $T$ is time-reversal)}, which raised a debate on whether a weak Coulomb interaction is sufficient to destroy the higher-order topological insulators~\cite{HOTIC1,HOTIC2,HOTIC3}. Therefore, how to characterize higher-order topological phases in the presence of interaction still remains an open question.

In this paper, we study the topological properties and stability of higher-order topological insulators under interaction. We focus on the $C_4T$-protected three dimensional second order topological insulator with interaction and show that its topological index can be computed in a gauge-independent way through eigenstates of the inverse Green's function at zero frequency, which is a generalization of the approach in Ref. \onlinecite{Wang2012inv,Wang2012int}. Furthermore, if there is an $S_4$ rotoinversion symmetry in addition to $C_4T$, the topological index $P_3$ in this interacting system can be determined by eigenvalues of $S_4$ at high symmetry momenta, similar to the Fu-Kane parity criterion~\cite{FuKane_inv}. {We demonstrate this method by computing the topological index of HOTIs with Coulomb interaction, where we obtain the Green's function from exact diagonalization (ED). We also discuss the realization of this higher-order topological phase in insulators with $C_4T$-preserving magnetic order. Finally, we investigate the influence of Coulomb interaction and boundary termination on the hinge states. We show that the gapless hinge states as the features of higher-order topology remain robust under weak Coulomb interaction that does not close the surface gap, and the hinge states can emerge even when boundary is smooth and without a sharp hinge.}


\section{Higher-order topological index with interaction}

\begin{figure}
\includegraphics[width=3.2 in]{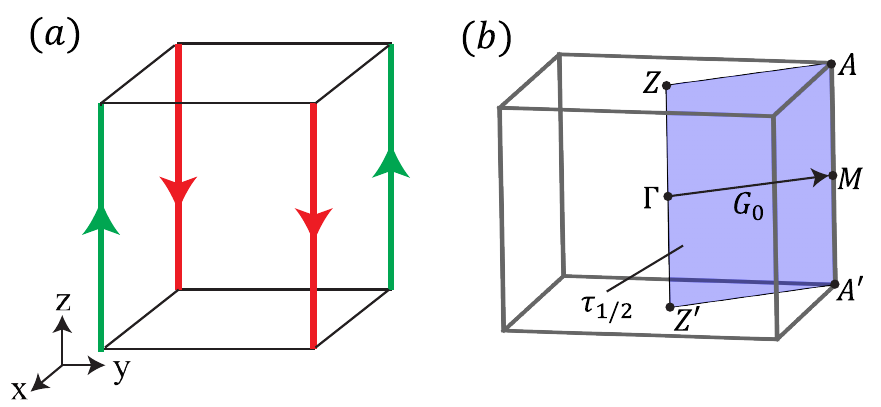}
\caption{ (a): Illustration of the chiral hinge modes in second-order topological insulator protected by $C_4T$ symmetry. (b): Brillouin zone of the second-order topological insulator. $\Gamma,Z,M,A$ are $S_4$-invariant momenta. The colored area represents the region $\tau_{1/2}$ for computing topological index.   }
\label{fig_hinge}
\end{figure}

Consider the 3D chiral second-order topological insulator protected by $C_4T$ symmetry proposed in Ref. \onlinecite{Bernevig_HOTI}. The higher-order topological feature is characterized by the chiral hinge states propagating in alternating directions at hinges parallel to the fourfold rotational axis, as shown in Fig.\ref{fig_hinge}(a). In the non-interacting limit, a tight-binding model for this second-order topological insulator is given by
\bea
H_{0}(\mathbf{k})&=& {\left[M+\sum_{i=x,y,z} t_{i} \cos \left( k_{i}a\right)\right] \tau_{z} \sigma_{0}+\sum_{i=x,y,z} \Delta_{i} \sin \left( k_{i}a\right) } \nonumber\\
&& \times \tau_{x} \sigma_{i}+\lambda_1 \sin (k_z a) \tau_y\sigma_0+\lambda_2\tau_x\sigma_0 \nonumber\\
&&+\Delta_{2}\left[\cos \left( k_{x}a\right)-\cos \left( k_{y}a\right)\right] \tau_{y} \sigma_{0}.
\label{HOTImodel}
\eea
{Here $\tau$ and $\sigma$ refer to the orbital and spin spaces respectively.} The fourfold rotation operator is $C_4=\tau_0e^{-i\frac{\pi}{4}\sigma_z}$ and time-reversal operator is $T=-i\tau_0\sigma_y K$, where $K$ is complex conjugation. Without the $\Delta_2$ term, the system is a first-order $\mathbb{Z}_2$ topological insulator with symmetries $C_4$ and $T$. The $\Delta_2$ term breaks $C_4$ and $T$ separately but preserves the product $C_4T$, which opens a surface gap and drives the system into a second-order topological insulator protected by $C_4T$ symmetry. The topological index for this system is the magneto-electric polarization $P_3$, which is quantized by $C_4T$ symmetry to $0$ or $\frac{1}{2}$ with a $\mathbb{Z}_2$ classification. {In the non-interacting limit, $P_3$ can be computed via the Bloch wave functions which are eigenstates of Eq.\eqref{HOTImodel}. When electron interactions are taken into account which goes beyond Eq.\eqref{HOTImodel}, the Bloch wave functions are no longer appropriate for computing the topological index.}

For an interacting system one can focus on the Green's function instead. For a $N$-band interacting system the Matsubara Green's function is an $N\times N$ matrix $G(i\omega,\mathbf k)=\left(i\omega-h(\mathbf k)-\Sigma(i\omega,\mathbf k)\right)^{-1}$, where $h(\mathbf k)$ is the non-interacting Hamiltonian matrix and $\Sigma(i\omega,\mathbf k)$ is the self-energy. We take the zero temperature limit in which the Matsubara frequency $\omega$ can take continuous values. The Lehmann representation requires $G^\dagger(-i\omega,\mathbf k)=G(i\omega,\mathbf k)$, which implies the Green's function (and its inverse) at zero frequency is a Hermitian matrix with real eigenvalues. $G^{-1}(0,\mathbf k)$ can be diagonalized as
\be
G^{-1}(0,\mathbf k)\ket{g_n(\mathbf k)}=\lambda_n(\mathbf k)\ket{g_n(\mathbf k)}.
\label{Geigen}
\ee
In the non-interacting limit $-G^{-1}(0,\mathbf k)$ is the Hamiltonian $h(\mathbf k)$ of an insulator with a band gap such that all eigenvalues of $h(\mathbf k)$ are nonzero. We assume that the interaction does not close the gap, hence in the interacting system all $\lambda_n(\mathbf k)$ are real and nonzero for every $\mathbf k$ as well. Denote the number of positive $\lambda_n$ at each momentum by $N_+$. {The magneto-electric polarization $P_3$ in general can be written as an integral that involves the Green's function and its inverse over the whole momentum and frequency space}~\cite{Qi2008}. Ref. \onlinecite{Wang2012int} shows that when the system has a unique ground state and a finite gap, $P_3$ can be simplified to involve only the eigenstates of the inverse Green's functions at zero frequency:
\begin{equation}
\begin{aligned}
P_{3} \left.= \int \frac{d^{3} \mathbf k}{8 \pi^{2}}  \epsilon^{i j k} \operatorname{Tr}\left\{\left[\partial_{i} \mathcal{A}_{j}(\mathbf k)+\frac{2}{3} i \mathcal{A}_{i}(\mathbf k) \mathcal{A}_{j}(\mathbf k)\right)\right] \mathcal{A}_{k}(\mathbf k)\right\}.
\label{P3int}
\end{aligned}
\end{equation}
Here $\mathcal{A}_{j}(\mathbf k)$ is an $N_+\times N_+$ matrix defined from the eigenstates of $G^{-1}(0,\mathbf k)$ with positive eigenvalues:
\be
\left[\mathcal{A}_{j}(\mathbf k)\right]_{mn}=-i\bra{g_m(\mathbf k)}\frac{\partial}{\partial k_j}\ket{g_n(\mathbf k)},\ \ \lambda_m(\mathbf k),\lambda_n(\mathbf k)>0.
\label{Aint}
\ee
Although Eq.\eqref{Aint} is similar to the non-Abelian Berry connection in the non-interacting systems, the physical meaning is very different, because Eq.\eqref{Aint} involves the eigenstates of inverse Green's function at zero frequency rather than Bloch wave functions. Eq.\eqref{P3int} is well-defined for interacting systems, but the direct computation from Eq.\eqref{P3int} is not practical due to the requirement of a global smooth gauge in $\mathcal{A}_{j}(\mathbf k)$. 

The symmetries in higher-order topological insulator can further simplify Eq.\eqref{P3int} so that a global smooth gauge is no longer needed. $C_4T$ symmetry requires
\be
(C_4T) G^{-1}(i\omega,\mathbf k)(C_4T)^{-1}=G^{-1}(-i\omega,C_4T\mathbf k),
\ee
where $C_4T\mathbf k=(k_y,-k_x,-k_z)$. This implies $C_4T\ket{g_n(\mathbf k)}$ is an eigenstate of $G^{-1}(0,C_4T\mathbf k)$ with the same energy. Therefore, the $C_4T\ket{g_n(\mathbf k)}$ can be expanded as $C_4T\ket{g_n(\mathbf k)}=\sum_m \ket{g_m(C_4T\mathbf k)}B(\mathbf k)_{mn}$ and the sewing matrix $B(\mathbf k)$ is unitary:
\be
B(\mathbf k)_{mn}=\bra{g_m(C_4T\mathbf k)}C_4T\ket{g_n(\mathbf k)}.
\label{Bsew}
\ee
The similarity between Eqs.\eqref{P3int}-\eqref{Bsew} and their Bloch counterpart in the non-interacting limit implies $P_3$ can be written as the wrapping number of sewing matrix $B(\mathbf k)$~\cite{Bernevig_HOTI}:
\begin{equation}
2 P_{3}=-\frac{1}{24 \pi^{2}} \int d^{3} \mathbf{k} \epsilon^{i j k} \operatorname{Tr}\left[\left(B \partial_{i} B^{\dagger}\right)\left(B \partial_{j} B^{\dagger}\right)\left(B \partial_{k} B^{\dagger}\right)\right].
\label{eqwrap}
\end{equation}
Using the degree counting method in Ref. \onlinecite{Li2020pf}, Eq.\eqref{eqwrap} can be reduced to a Pfaffian formula:
\bea
2P_3&=&\frac{1}{2\pi i}\oint_{\partial\tau_{1/2}} d\mathbf{k} \cdot \mathbf{\nabla} \log \operatorname{Pf}[M(\mathbf{k})],
\label{Pfmain} \\
M(\mathbf{k})_{m n}&=&\bra{ g_{m}(\mathbf{k})}\frac{C_{4} T+C_{4}^{-1} T}{\sqrt{2}}\ket{ g_{n}(\mathbf{k})}.
\eea
Here $\operatorname{Pf}$ denotes Pfaffian, which is only defined for anti-symmetric matrices. The anti-symmetric property of matrix $M$ is guaranteed by $(C_4T)^4=-1$, as shown in the appendix. $\partial\tau_{1/2}$ is the boundary of the region $\tau_{1/2}$ in Fig.\ref{fig_hinge}(b). Because the integral along $ZA$ and $Z'A'$ cancel each other by periodicity, the integral only involves $ZZ'$ and $AA'$. Importantly, to evaluate Eq.\eqref{Pfmain} one needs to make a gauge choice such that $\det[B(\mathbf k)]$ is smooth in $\tau_{1/2}$. 


Starting from the Pfaffian formula Eq.\eqref{Pfmain}, a gauge-independent method can be developed to compute the topological index $P_3$, which only involves the inverse Green's function at momenta inside $\tau_{1/2}$~\cite{LiTIC2T}. Define a gauge-invariant quantity $\mathcal R$ on a straight line connecting $\mathbf k_a,\mathbf k_b$ in momentum space:
\bea
\mathcal R(\mathbf k_a,\mathbf k_b)&=&\frac{\operatorname{Pf}[M(\mathbf k_b)]}{\operatorname{Pf}[M(\mathbf k_a)]}\det[ W(\mathbf k_a,\mathbf k_b) ], \label{gwilson}\\
W_{mn}(\mathbf k_a,\mathbf k_b)&=&\langle g_m(\mathbf k_a)| \prod_{\mathbf k_i\in \overline{\mathbf k_a \mathbf k_b}}^{\mathbf k_a\leftarrow \mathbf k_b} P_{\mathbf k_i} | g_n(\mathbf k_b)\rangle.
\eea
Here $P_{\mathbf k}=\sum_{m} \ket{g_m(\mathbf k)}\bra{g_m(\mathbf k)}$ is the projection to the space spanned by eigenstates of inverse Green's function with positive eigenvalues, and $\overline{\mathbf k_a \mathbf k_b}$ denotes the straight line connecting $\mathbf k_a$ and $\mathbf k_b$. {The arrow $\mathbf k_a\leftarrow \mathbf k_b$ denotes the direction of path-ordered product which puts momentum points close to $\mathbf k_b$ to the right.} The path-ordered product in $W$ is similar to the Wilson loop but it is defined in interacting systems. Let $\mathbf G_0$ be the vector connecting $\Gamma$ and $M$ points in the Brillouin zone. As shown in the appendix, $\mathcal R(\mathbf k_a,\mathbf k_b)$ is invariant under gauge transformation $|g_n(\mathbf k)\rangle\rightarrow\sum_m |g_m(\mathbf k)\rangle U_{mn}(\mathbf k)$ and the Pfaffian formula in Eq.\eqref{Pfmain} can be computed by the following integral along the straight line connecting $Z'$ and $Z$~\cite{LiTIC2T}:
\be
2P_3=\frac{1}{2\pi i} \int_{Z'}^Z   d\mathbf k \cdot \bold{\nabla} \log \mathcal R(\mathbf k,\mathbf k+\mathbf G_0).
\label{P3ginv}
\ee
{Eq.\eqref{P3ginv} can be understood as follows. It can be shown that $\mathcal R(\mathbf k_a,\mathbf k_b)$ reduces to the ratio of the Pfaffian of $M$ matrix between $\mathbf k_b$ and $\mathbf k_a$ under a suitable gauge choice~\cite{LiTIC2T}. Then Eq.\eqref{P3ginv} measures the winding of the phase of Pfaffian along $ZZ'$ and $AA'$, which is equivalent to Eq.\eqref{Pfmain}. A more rigorous proof is shown in the appendix. In practical computation, the path ordered product in $W$ can be evaluated at discrete momentum points similar to the computation of Wilson loop because the phase of $W$ is insensitive to discretization of momentum points.} Due to the gauge-invariance of $\mathcal R$, the evaluation of Eq.\eqref{P3ginv} does not require a smooth gauge, and it can be computed in any gauge obtained directly from diagonalizing the inverse Green's function. Therefore Eq.\eqref{P3ginv} provides an efficient gauge-independent method for computing topological index $P_3$ in interacting systems. 



If the system has a fourfold rotoinversion symmetry $S_4=C_4I$ in addition to $C_4T$ symmetry where $I$ denotes space inversion operator, the eigenstates $\ket{g_n(\mathbf k)}$ of $G^{-1}(0,\mathbf k)$ at $S_4$-invariant momentum $\mathbf k$ are also simultaneous eigenstates of $S_4$, leading to
\be
S_4\ket{g_n(\mathbf k)}=s_n(\mathbf k)\ket{g_n(\mathbf k)},\ \mathbf k\in K^4.
\ee
Here $K^4$ is the set of four high symmetry momenta that are invariant under $S_4$ in the 3D Brillouin zone, as shown in Fig.\ref{fig_hinge}(b). Following Ref. \onlinecite{Li2020pf}, in the presence of $S_4$ symmetry Eq.\eqref{Pfmain} can be simplified to the product of $(s_n+s_n^{-1})/\sqrt{2}=\pm 1$ at high symmetry momenta:
\be
(-1)^{2P_3}=\prod_{\mathbf k\in K^4}\prod_n \frac{s_n(\mathbf k)+s_n(\mathbf k)^{-1}}{\sqrt{2}}.
\label{P3S4}
\ee
Here the product of $n$ is over the eigenstates of $G^{-1}(0,\mathbf k)$ with positive eigenvalues, and only one state in each Kramers pair is taken in the product. Eq.\eqref{P3S4} shows that in the presence of finite interaction, although Bloch wave functions can no longer be applied to compute topological index, an alternative route is provided by the eigenstates of inverse Green's function at zero frequency such that topological indices can still be extracted from eigenvalues of symmetry operators.

\section{Green's function method in other higher-order topological phases}

{The Green's function method can also be applied to some other types of higher-order topological insulators with interaction. For example, the 3D helical HOTI proposed in Ref.\onlinecite{Bernevig_HOTI} is protected by time-reversal symmetry and a pair of perpendicular mirror symmetries $M_{xy}$ and $M_{x\overline{y}}$. The hinge states appear at the mirror-invariant hinges when the corresponding mirror Chern number $C_m$ is a nonzero even number. The two mirror symmetries constitutes a $\mathbb{Z}\times\mathbb{Z}$ classification and the topological index is $C_m/2$. When there is finite interaction, the Bloch wave functions are not available to compute the mirror Chern number, but it can still be computed from the eigenstates of the inverse Green's function obtained in Eq.\eqref{Geigen}. Because the inverse Green's function still preserves mirror symmetry, one can select out the eigenstate $|g(\mathbf k)\rangle$ of inverse Green's function that is simultaneous eigenstate of the mirror symmetry. Then the effective Berry connection is given by $\mathcal{A}_j(\mathbf k)=-i\langle g(\mathbf k)|\partial_{k_j}|g(\mathbf k)\rangle$ and the mirror Chern number can be computed by $C_m=\int d^2k (\partial_{k_x}\mathcal{A}_y(\mathbf k)-\partial_{k_y}\mathcal{A}_x(\mathbf k))$, where the integral is inside the mirror-symmetric plane. This is another example to use Green's function to compute the topological index for HOTIs with interaction.}

Higher-order topological phases can also be realized in superconductors. Our method based on Green's function is applicable to higher-order topological superconductors as well. The Green's function $\mathcal G$ for superconductors includes both the particle-hole and particle-particle channels~\cite{Wang2012GSC}:
\begin{equation}
\mathcal{G}(i \omega, \mathbf k)=\left(\begin{array}{ll}
G_{A}(i \omega, \mathbf k) & G_{B}(i \omega, \mathbf k) \\
G_{C}(i \omega, \mathbf k) & G_{D}(i \omega, \mathbf k)
\end{array}\right),
\end{equation}
where
\bea
\left(G_{A}\right)_{\alpha \beta}(i \omega, \mathbf k) &=&-\int_{0}^{\beta} d \tau e^{i \omega \tau}\left\langle T_{\tau} c_{\mathbf k \alpha}(\tau) c_{\mathbf k \beta}^{\dagger}(0)\right\rangle, \nonumber\\
\left(G_{B}\right)_{\alpha \beta}(i \omega, \mathbf k) &=&-\int_{0}^{\beta} d \tau e^{i \omega \tau}\left\langle T_{\tau} c_{\mathbf k \alpha}(\tau) c_{-\mathbf k \beta}(0)\right\rangle, \nonumber\\
\left(G_{C}\right)_{\alpha \beta}(i \omega, \mathbf k) &=&-\int_{0}^{\beta} d \tau e^{i \omega \tau}\left\langle T_{\tau} c_{-\mathbf k \alpha}^{\dagger}(\tau) c_{\mathbf k \beta}^{\dagger}(0)\right\rangle, \nonumber\\
\left(G_{D}\right)_{\alpha \beta}(i \omega, \mathbf k) &=&-\int_{0}^{\beta} d \tau e^{i \omega \tau}\left\langle T_{\tau} c_{-\mathbf k \alpha}^{\dagger}(\tau) c_{-\mathbf k \beta}(0)\right\rangle. \nonumber\\
\eea
Then one can obtain the eigenstates of the inverse Green's function at zero frequency
\be
\mathcal G^{-1}(0,\mathbf k)\ket{g_n(\mathbf k)}=\lambda_n(\mathbf k)\ket{g_n(\mathbf k)},
\ee
and $\ket{g_n(\mathbf k)}$ can be utilized to compute the topological index. For interacting second-order topological superconductor protected by $C_4T$ symmetry, the topological index can be computed via Eq.\eqref{P3ginv}, and with an additional $S_4$ symmetry it can be computed via Eq.\eqref{P3S4}.

The $C_4T$-symmetric second-order topological superconductor can be realized by $p+id$ pairing as shown in Ref. \onlinecite{Wang2018SC}:
\bea
\hat H&=&\sum_{\mathbf k}c^\dagger_{\mathbf k}(\frac{k^2}{2m}-\mu)c_{\mathbf k}+\Delta_p c^T_{\mathbf k}(\mathbf k\cdot \sigma)i\sigma_y c_{-\mathbf k} \nonumber\\
&&+i\Delta_d c^T_{\mathbf k}(k_x^2-k_y^2)i\sigma_y c_{-\mathbf k} + h.c.
\eea
The lattice-regularized BdG Hamiltonian in the Nambu basis $\Psi_{\mathbf k}=(c_{\mathbf k\uparrow},c_{\mathbf k\downarrow},c^\dagger_{-\mathbf k\uparrow},c^\dagger_{-\mathbf k\downarrow})^T$ is given by:
\bea
\hat H&=&\sum_{\mathbf k}\Psi^\dagger_{\mathbf k}H_{\mathbf k}\Psi_{\mathbf k}, \nonumber \\
H_{\mathbf k}&=&\left(2t\cos (k_x a)+2t\cos (k_y a)+2t_z\cos (k_z c)-\mu\right)\nonumber\\
&&\times\tau_z\sigma_0 -\Delta_p\sin (k_x a)\tau_x\sigma_z \nonumber\\
&&+\Delta_p \sin (k_y a)\tau_y\sigma_0+\Delta_p\sin (k_z c)\tau_x\sigma_x  \nonumber\\
&&+\Delta_d(\cos(k_xa)-\cos(k_ya))\tau_x\sigma_y.
\label{HOSC}
\eea
{Here $\tau$ is the particle-hole space and $\sigma$ is the spin space.}  Without the $d$-wave term $\Delta_d$, this Hamiltonian describes a first-order topological superconductor with $p$-wave pairing. It has time-reversal symmetry $T=-i\tau_0\sigma_yK$, fourfold rotation symmetry $C_4=\text{diag}\{e^{-\frac{i\pi}{4}},e^{\frac{i\pi}{4}},e^{\frac{i\pi}{4}},e^{-\frac{i\pi}{4}}\}=2^{-1/2}(\tau_0\sigma_0-\tau_z\sigma_z)$ and an effective "inversion" symmetry $I=\tau_z\sigma_0$ that satisfies $IH_{\mathbf k}I^{-1}=H_{-\mathbf k}$. The $d$-wave term flips sign under $C_4,T,I$ symmetries separately but preserves the products $C_4T$ and $S_4=C_4I$. The $d$-wave term also anti-commutes with the other terms in the Hamiltonian, hence it can open a surface gap to drive the system to a higher-order topological superconductor with chiral Majorana hinge states. Note that a $\pi/2$ phase difference between the $p$- and $d$-wave pairing is needed, otherwise the $d$-wave term will be $\tau_y\sigma_y$ instead which cannot open a surface gap. {If we go beyond the mean-field level and take into account the quasiparticle interactions that are not included in Hamiltonian Eq.\eqref{HOSC}, the topological index cannot be computed from the wave functions obtained by diagonalization of the BdG Hamiltonian. Instead, it can be computed from the eigenstates of the inverse Green's function $\mathcal G$ with the help of Eq.\eqref{P3ginv} or \eqref{P3S4}, similar to the case of higher-order topological insulators.}

\section{Numerical demonstration of the Green's function approach}
\label{Sec_ED}


We demonstrate the implementation of the Green's function method by solving for the eigenstates of an interacting higher-order topological insulator via exact diagonalization (ED). The non-interacting part of the Hamiltonian is the same as Eq.\eqref{HOTImodel}, which describes a four-band 3D insulator with $C_4T$ symmetry on a tetragonal lattice with two orbitals $A,B$ in each unit cell. The conduction and valence bands are separated by an energy gap $\Delta$. Denote the electron creation operator in the orbital space by $c^\dagger_{\mathbf k,\ell s}$ where $\ell=A,B$ and $s=\uparrow,\downarrow$. Then the creation operator for the n-th single particle band $\psi^\dagger_{\mathbf k,n}$ is related by $c^\dagger_{\mathbf k,\ell s}=\sum_n \psi^\dagger_{\mathbf k,n}[u_n(\mathbf k)]^*_{\ell s}$, where $u_n(\mathbf k)$ is a four-component column vector for the n-th band of the single particle Hamiltonian $H_0$. Consider a repulsive Hubbard-like interaction:
\bea
H_{int}&=&\frac{U}{N_{cell}}\sum_{\mathbf k,\mathbf k',\mathbf q,\ell}c^\dagger_{\mathbf k+\mathbf q,\ell\uparrow}c_{\mathbf k,\ell\uparrow} c^\dagger_{\mathbf k'-\mathbf q,\ell\downarrow}c_{\mathbf k',\ell\downarrow}.
\label{Hint}
\eea
Here $N_{cell}$ is the number of unit cells. For simplicity we assume the interaction is local in momentum $k_z$. To perform exact diagonalization, we choose the basis of many-body states to have the form $|\Psi^{basis}\rangle=\prod_{i=1}^N\psi^\dagger_{\mathbf k_i,n_i}|0\rangle$, where $|0\rangle$ is the vacuum state and $N$ is the total number of electrons, which equals to the number of states in the valence bands $N=2N_{cell}$. A generic many-body state $|\Psi\rangle$ is a superposition of different basis states. For each basis state $|\Psi^{basis}\rangle$, denote the number of electrons in the conduction bands by $N_c$. Then $N_c=0$ corresponds to a unique state with all electrons filling up the valence bands, which is the ground state of single particle Hamiltonian. There are $N^2$ basis states with $N_c=1$, corresponding to exciton excitation obtained by moving an electron from the fully filled valence bands to conduction bands. Due to the energy gap $\Delta$, in the non-interacting limit the energy of the states with larger $N_c$ is higher than the ground state by at least $N_c\Delta$. Interaction can mix many-body states with different $N_c$ so that $N_c$ is no longer a good quantum number. However, for weak interaction $U\ll \Delta$ the many-body ground state should still mainly consist of states with small $N_c$. Therefore, in order to determine the ground state and Green's function under weak interaction, it is sufficient to restrict the Hilbert space to many-body states with small $N_c$.

The computation of topological index $P_3$ involves $\mathcal R(\mathbf k,\mathbf k+\mathbf G_0)$ in Eq.\eqref{P3ginv} where $\mathbf k=(0,0,k_z)$ and $\mathbf G_0=(\frac{\pi}{a},\frac{\pi}{a},0)$. Denote the system size along $x$ and $y$ directions as $L\times L$. The minimal system size that can preserve the $C_4T$ symmetry and make $\mathcal R(\mathbf k,\mathbf k+\mathbf G_0)$ well-defined is $L=4$. To keep the size of Hilbert space manageable, we only consider many-body states with $N_c=0,1,2$ and focus on the weak and intermediate interaction region. $\mathcal R(\mathbf k,\mathbf k+\mathbf G_0)$ depends only on $k_z$, hence we treat each $k_z$ independently and for a given $k_z$ the system can be treated as quasi-2D with $N=32$ electrons which are enough to fill up two valence bands in a lattice of size $4\times 4$. Due to the translational symmetry, each many-body state has a well-defined total momentum $\mathbf k=\frac{k_1}{L}\mathbf G_x+\frac{k_2}{L}\mathbf G_y$, where $\mathbf G_x$ and $\mathbf G_y$ are reciprocal lattice vectors along $x$ and $y$ respectively. The many-body Hamiltonian is block-diagonal and the energy spectrum can be resolved for each distinct total momentum.

\begin{figure}
\includegraphics[width=3.4 in]{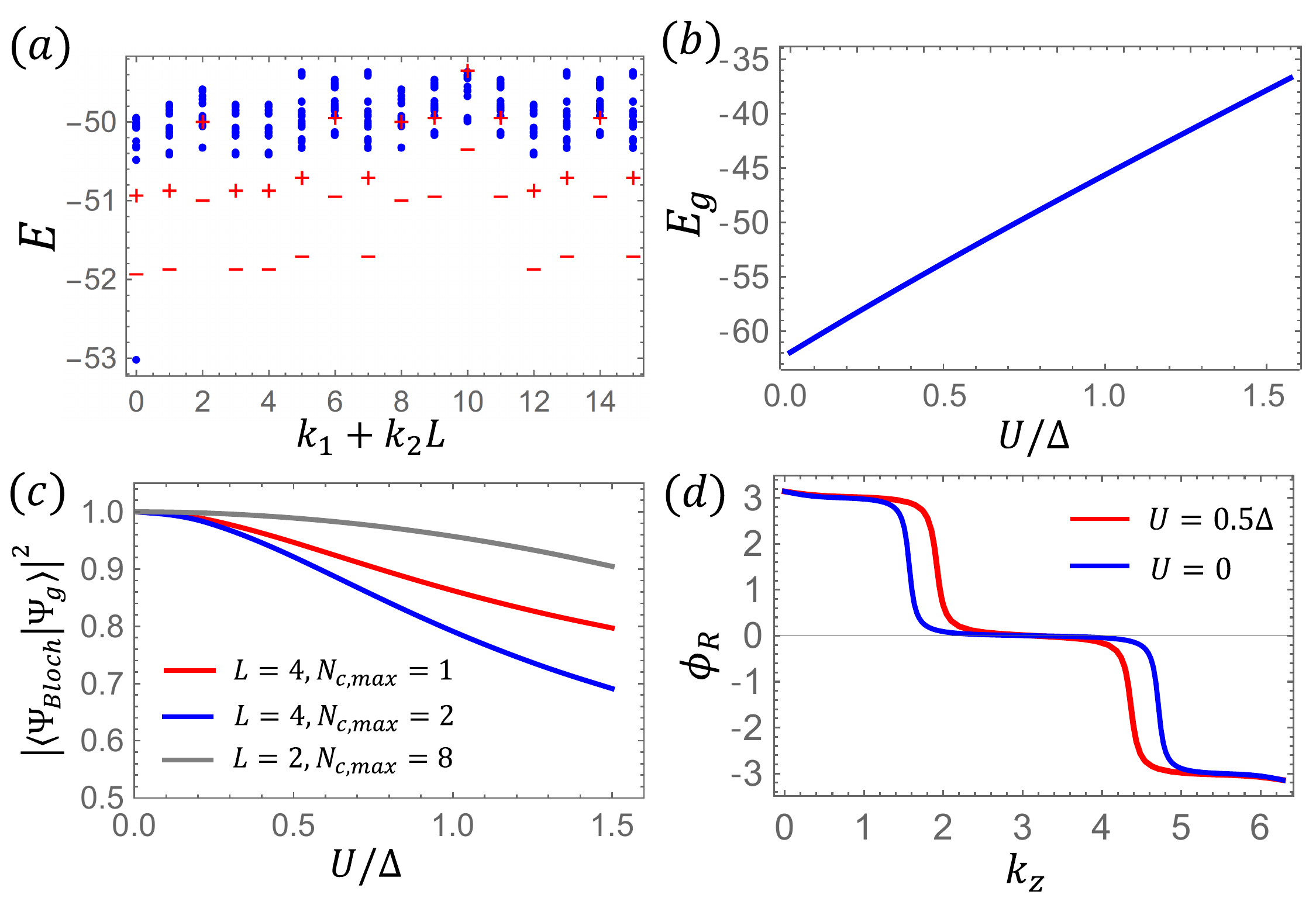}
\caption{ (a): Energy spectrum obtained from exact diagonalization for system size $L=4$ with finite interaction $U=1$. Each many-body state is resolved by total momentum $\mathbf k=\frac{k_1}{L}\mathbf G_x+\frac{k_2}{L}\mathbf G_y$. The blue dots represent the spectrum of half-filled system with $N=32$ particles. The red $\pm$ signs represent the single particle/hole excitation obtained from systems with $N\pm 1$ particles. (b): Ground state energy as a function of interaction strength. $\Delta=1.8$ is the band gap in the non-interacting limit. (c): Overlap between the interacting ground state and the non-interacting ground state. (d): The phase $\phi_R$ of $\mathcal R(\mathbf k,\mathbf k+\mathbf G_0)$ as a function of $k_z$. The winding of $\phi_R$ with $k_z$ indicates nontrivial topological index.  }
\label{fig_ED}
\end{figure}

The spectrum obtained from ED for finite interaction $U=1$ with parameters $M=-2,t_i=1,\Delta_i=1 (i=x,y,z),\Delta_2=1,\lambda_1=0.3,\lambda_2=0.2$ at $k_z=0$ is shown as the blue dots in Fig.\ref{fig_ED}(a). The ground state $|\Psi_g\rangle$ has zero total momentum, and is separated from the higher-energy states by a gap $\tilde\Delta=2.6$. If there is no interaction, this gap is the same as the single particle band gap $\Delta=1.8$ and the ground state is the direct product of all Bloch states in the occupied valence band $|\Psi_{Bloch}\rangle=\prod_{\mathbf k_i,n_i\in occ}\psi^\dagger_{\mathbf k_i,n_i}|0\rangle$. As interaction strength increases, the ground state energy increases as shown in Fig.\ref{fig_ED}(b),  consistent with the expectation of a repulsive interaction. Furthermore, the ground state $|\Psi_g\rangle$ also deviates from the direct product state $|\Psi_{Bloch}\rangle$ when there is finite interaction. Fig.\ref{fig_ED}(c) shows the overlap $|\langle\Psi_{Bloch}|\Psi_g\rangle|^2$ as a function of interaction strength. The blue curve represents the result from ED by taking into account the many-body states with $N_c=0,1,2\ (N_{c,\rm{max}}=2)$ and the red curve represents those by taking only $N_c=0,1\ (N_{c,\rm{max}}=1)$. As interaction increases, the ground state overlap decreases, and it reaches $\sim$90\% when $U=0.5\Delta$. At the same time, the approximation in ED that only takes many-body states with small $N_c$ becomes less accurate when interaction strength increases, which is indicated by the difference between the results from $N_{c,\rm{max}}=2$ and $N_{c,\rm{max}}=1$. When $U\gtrsim\Delta$ there is sizable difference between the two curves, indicating the many-body states with larger $N_c$ need to be taken into account. We checked in a smaller system size $L=2$ that including states with larger $N_c$ does not qualitatively change the results. We performed full ED computation with all $N_c$ included for $L=2$ as shown in the grey curve in Fig.\ref{fig_ED}(c). At large $U$ the exact ground state is still non-degenerate and separated from the other states by an energy gap, and it has a sizable overlap with $|\Psi_{Bloch}\rangle$. Therefore, for the $L=4$ system the approximation of taking $N_{c,\rm{max}}=2$ still gives a valid description of the ground state if we focus on the weak interaction region.

The Green's function can be computed via Lehmann representation:
\bea
G_{\alpha \beta}(0, \mathbf k)&=& \sum_{m}\left[\frac{\langle \Psi_g|c_{\mathbf k \alpha}| \Psi_m\rangle\langle \Psi_m|c_{\mathbf k \beta}^{\dagger}| \Psi_g\rangle}{-\left(E_{m}-E_{g}\right)}\right.\nonumber\\
&&\left.+\frac{\langle \Psi_m|c_{\mathbf k \alpha}| \Psi_g\rangle\langle \Psi_g|c_{\mathbf k \beta}^{\dagger}| \Psi_m\rangle}{\left(E_{m}-E_{g}\right)}\right].
\label{Gleh}
\eea
Here $\alpha,\beta$ represent combined indices for orbital and spin. $|\Psi_g\rangle$ is the ground state with $N$ electrons and $\ket{\Psi_m}$ represents excited state with $N\pm 1$ electrons. Therefore, calculating the Green's function for a $N$-particle system also requires ED computation for systems with $N\pm 1$ particles, which correspond to single-particle ($N+1$) and single-hole ($N-1$) excitation above the $N$-particle ground state. The spectrum for $N\pm 1$ particles is also shown in Fig.\ref{fig_ED}(a) as the red signs, where only the lowest energy state is shown for each momentum.

We compute the Green's function for systems with size $L=4$ by Eq.\eqref{Gleh}. An effective Hamiltonian can be defined by $H_{\rm{eff}}(\mathbf k)=-G(0,\mathbf k)^{-1}$. In the non-interacting limit $H_{\rm{eff}}$ is the same as $H_0$, and with finite interaction $H_{\rm{eff}}$ changes but it still preserves the symmetries. As long as $H_{\rm{eff}}$ remains gapped, the quantity $\mathcal R(\mathbf k,\mathbf k+\mathbf G_0)$ is well-defined and can be used to compute the topological index. We calculate $\mathcal R(\mathbf k,\mathbf k+\mathbf G_0)$ where $\mathbf k=(0,0,k_z)$ using the eigenvectors of the inverse Green's function. Denote the phase of $\mathcal R(\mathbf k,\mathbf k+\mathbf G_0)$ as $\phi_{R}$. The evolution of $\phi_{R}$ as a function of $k_z$ for $U=0.5\Delta$ and $U=0$ systems are shown in Fig.\ref{fig_ED}(d). In both cases the phase $\phi_{R}$ shows a winding $2\pi$ as $k_z$ increases, hence by Eq.\eqref{P3ginv} the topological index $P_3$ is nontrivial for both systems. If interaction becomes stronger $U\gtrsim\Delta$, the single-hole excitation energy can drop below the ground state energy, and the eigenvalues of the inverse Green's function will cross zero. Then the effective Hamiltonian is no longer gapped, which leads to a topological transition out of the HOTI phase. Note that for the finite interaction $U=0.5\Delta$ in Fig.\ref{fig_ED}(c,d) the effective Hamiltonian $H_{\rm{eff}}$ is still gapped and the topological index is well-defined. However, the many-body ground state is no longer a simple direct product state and the non-interacting formalism for topological indices are no longer applicable, but the Green's function method still remains valid. Therefore, the Green's function method is useful to identify the topological properties of interacting systems.

\section{Realization of higher-order topological phases}

The higher-order topological insulators protected by $C_4T$ symmetry discussed above can be generated by appropriate magnetic order that breaks time-reversal and fourfold rotation symmetries down to $C_4T$. Here we discuss some possible crystal structure and the corresponding magnetic order that can realize this phase.

\begin{figure}
\includegraphics[width=3.3 in]{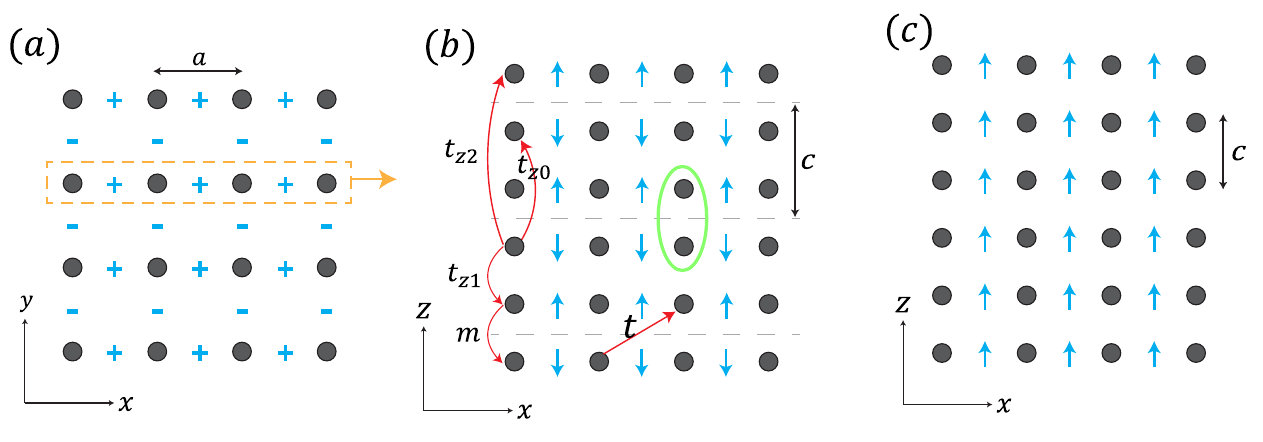}
\caption{ Top view (a) and front views (b),(c) of lattices that can realize the higher-order topological insulator. The active electrons are at black sites. The green circle in (b) denotes the two atoms in each unit cell. The blue arrows and signs represent local moments with magnetization. The lattice with combination (a),(b) has $C_4T$ and $S_4$ symmetry. The lattice with combination (a),(c) has $C_4T$ and inversion symmetry.   }
\label{fig_latticeH}
\end{figure}

Consider the lattice represented by the black sites in Fig.\ref{fig_latticeH}(a),(b), where (a) and (b) are the top view and front view of the three dimensional lattice respectively. Suppose there is one orbital and one electron per site, and the unit cell contains two sites denoted by the green circle. The blue arrows represent magnetization on a different type of atoms. Without the blue arrows, the black sites form a lattice with $C_4$ symmetry along $z$ direction, inversion symmetry at the center of unit cell and time-reversal symmetry. With appropriate hopping amplitudes and spin-orbit coupling, electrons at the black sites can form a first-order topological insulator. We denote the hopping parameters in Fig.\ref{fig_latticeH}(b) and choose the basis $(c_{A\uparrow},c_{A\downarrow},c_{B\uparrow},c_{B\downarrow})^T$ where $A,B$ represent the two sites in each unit cell. Up to an identity matrix that does not change the topology, the Hamiltonian is
\bea
H_{TI}(\mathbf k)&=&\left(2t\cos (k_x a)+2t\cos (k_y a)+2t_z\cos (k_z c)+m\right)\nonumber\\
&&\times\tau_x\sigma_0 +\lambda \sin (k_x a)\tau_z\sigma_x+\lambda \sin (k_y a)\tau_z\sigma_y \nonumber\\
&&+2t_z'\sin (k_z c)\tau_y\sigma_0.
\eea
{Here $\tau$ and $\sigma$ denote the sublattice and spin spaces respectively.} $t_z=(t_{z1}+t_{z2})/2,t_z'=(t_{z1}-t_{z2})/2$ and $\lambda$ is from spin-orbit coupling. This system has time-reversal symmetry $T=-i\tau_0\sigma_yK$, fourfold rotational symmetry $C_4=\tau_0e^{-i\frac{\pi}{4}\sigma_z}$ and inversion symmetry $I=\tau_x\sigma_0$. When $\max\{4|t|-2|t_z|,2|t_z|\}<|m|<4|t|+2|t_z|$ it is a first-order topological insulator protected by time-reversal symmetry. 

The presence of magnetization represented by the blue arrows breaks the symmetry into $C_4T$ and $S_4=C_4I$. {The hybridization between the states at the black and blue sites can modify the hopping amplitude between electrons on the black sites in a spin-dependent way. For example, for electrons hopping between two neighboring black sites, if the magnetization on the blue site in the middle of the hopping path is along $+z$ direction, then electrons with spin along $-z$ direction on one black site can hop to the middle blue site and then hop to the next black site, while for electrons on the black site with spin along $+z$ direction this hopping process mediated by the middle site is Pauli-blocked. Therefore, the presence of the magnetization in the hopping path can generate a spin-dependent hopping term proportional to $\sigma_z$. From the magnetic order in Fig.\ref{fig_latticeH}(a), the magnetization that surrounds electrons on black sites is positive along $\pm x$ direction and negative along $\pm y$ direction, and from Fig.\ref{fig_latticeH}(b) the magnetization around the two sublattices is opposite, this magnetic order generates a new term $(\cos(k_x a)-\cos(k_y a))\tau_z\sigma_z$ so that the Hamiltonian with magnetic order becomes}
\be
H_{HOTI}(\mathbf k)=H_{TI}(\mathbf k)+\Delta_2\left(\cos(k_x a)-\cos(k_y a)\right)\tau_z\sigma_z.
\label{HOTImag}
\ee
Eq.\eqref{HOTImag} is equivalent to the higher-order topological insulator in Eq.\eqref{HOTImodel} up to a unitary transformation. Therefore the magnetic order drives the system into a higher order topological insulator protected by $C_4T$ symmetry. The pattern of magnetization also indicates the system has an additional $S_4$ symmetry. 


The $C_4T$-symmetric higher-order topological insulator can also be realized in the lattice denoted by Fig.\ref{fig_latticeH}(a),(c). Here each unit cell has one black site and each site has one $s$-type and one $p_z$-type orbitals. Band inversion can occur between bands generated by the even- and odd-parity orbitals, leading to a first-order topological insulator. {The magnetic order modifies the hopping amplitude between electrons on black sites by adding hopping terms proportional to $\pm\sigma_z$ similar to the above analysis. These spin-dependent hopping terms break time-reversal symmetry. Because the magnetic order preserves $C_4T$ symmetry, the hopping terms it generates also preserve $C_4T$, which drives the system to $C_4T$-protected higher-order topological insulator.} Contrary to the magnetic order given by Fig.\ref{fig_latticeH}(a)(b), the lattice (a)(c) preserves inversion symmetry and breaks $S_4$ symmetry.

\begin{figure}
\includegraphics[width=3.3 in]{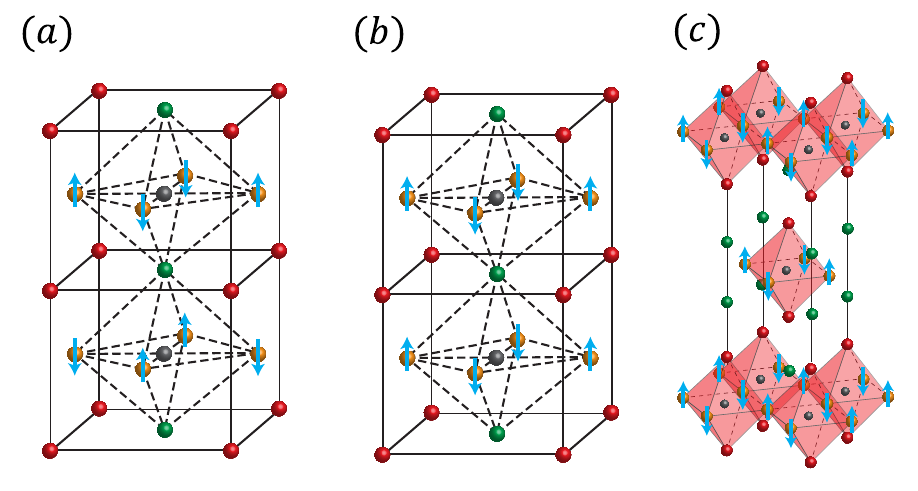}
\caption{ Crystal structures that can support the lattices in Fig.\ref{fig_latticeH}. Active electrons are from the black sites at the center, and the in-plane antiferromagnetic ordering denoted by blue arrows breaks time-reversal symmetry to drive the system to a $C_4T$-protected higher-order topological insulator.  }
\label{fig_structureH}
\end{figure}

The lattice in Fig.\ref{fig_latticeH}(a),(b) can be realized in the crystal structures in Fig.\ref{fig_structureH}(a). This structure can be viewed as a variation of antiperovskite structure ABX$_3$ with the top $X$ ion replaced by a different type of element, and the $c$ axis can acquire a lattice constant different from the $ab$ plane. If the bands close to Fermi level are from the B sites in the center, and a Neel order indicated by the blue arrows is developed surrounding the B sites, this structure can reproduce the physics in Fig.\ref{fig_latticeH}(a),(b) to realize a higher-order topological insulator protected by $C_4T$ symmetry. Similarly, the lattice in Fig.\ref{fig_latticeH}(a),(c) can be realized by the magnetic order in Fig.\ref{fig_structureH}(b). The active electrons can also sit on body-centered tetragonal Bravais lattice as in the anti-ruddlesden-popper structure in Fig.\ref{fig_structureH}(c). If the active electrons are on the red sites in the center, the in-plane antiferromagnetic ordering indicated by the blue arrows breaks time-reversal but preserves $C_4T$ and can drive the system into a higher-order topological insulator.

{The Coulomb interaction is ubiquitous in real materials. If Coulomb interaction is taken into account, the ground state is no longer a simple direct product of Bloch states but the higher-order topological features still persist for weak interaction. For the interacting system we can go through a procedure similar to Sec. \ref{Sec_ED} and use the eigenstates of inverse Green's function and the symmetry operators to compute the topological index via Eq.\eqref{P3ginv} or Eq.\eqref{P3S4}.}

\section{Stability of higher-order topological phases}

When the higher-order topological index for an interacting system is nontrivial, gapless hinge modes are expected to emerge from bulk-boundary correspondence. In this section we discuss the stability of these hinge modes against Coulomb interaction in the bulk and deformation at the boundary, and show that a finite surface gap is crucial to ensure the stability of these hinge modes.

\subsection{Stability against Coulomb interaction}

\begin{figure}
\includegraphics[width=3.4 in]{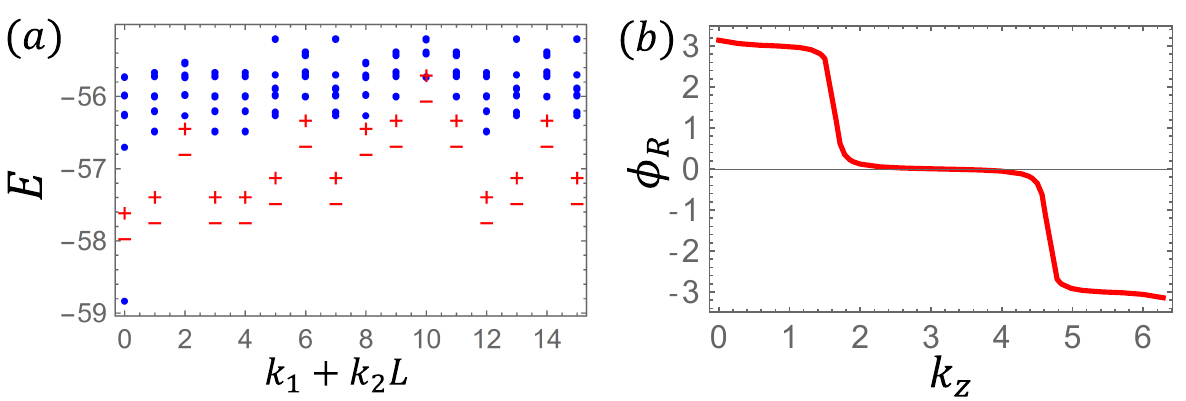}
\caption{ (a): The blue dots represent the energy spectrum for half-filled HOTI with long-range interaction $U=0.2,V_0=0.01$. The parameters are $M=-2,t_i=1,\Delta_i=1 (i=x,y,z),\Delta_2=1,\lambda_1=0.3,\lambda_2=0.2,k_z=0$. The red signs represent the single-particle and single-hole excitations with $N\pm 1$ particles. (b): Evolution of the phase $\phi_R$ of $\mathcal R(\mathbf k,\mathbf k+\mathbf G_0)$ as a function of $k_z$ using the same parameters as (a).   }
\label{fig_EDL}
\end{figure}

{Higher-order topological insulators have an energy gap in both the bulk and surface. If perturbations are added to the system, it is generally expected that as long as the perturbations are not strong enough to make the gap vanish, the higher-order topological phase should be robust. However, there is a recent debate in the literature on whether this HOTI phase is stable under Coulomb interaction~\cite{HOTIC1,HOTIC3,HOTIC2}. In particular, Refs.~\onlinecite{HOTIC1,HOTIC3} use the renormalization group and conclude that an infinitesimal long-ranged Coulomb interaction can destroy the HOTI phase. On the other hand, Ref.~\onlinecite{HOTIC2} argues that it is stable against a weak Coulomb interaction due to a different criterion for the stability. In order to test these claims, we perform a ED computation by adding a weak long-ranged interaction $H_V$ as well as the Hubbard-like interaction in Eq.\eqref{Hint} to the free HOTI Hamiltonian in Eq.\eqref{HOTImodel}:
\be
H_V=\sum_{i\ne j,\alpha\beta}V_{ij}c^\dagger_{i\alpha}c_{i\alpha}c^\dagger_{j\beta}c_{j\beta},\ \ V_{ij}=\frac{V_0}{r_{ij}}.
\ee
Here $i,j$ label the location of unit cell, $r_{ij}$ is the distance between $i$ and $j$ in units of nearest neighbor distance, and $\alpha,\beta$ are combined orbital and spin indices. We computed the Green's function and the phase $\phi_R$ of $\mathcal R(\mathbf k,\mathbf k+\mathbf G_0)$ under weak interaction $U=0.2$ and $V_0=0.01$ and find that $\phi_R$ still shows a winding with $k_z$ as shown in Fig.\ref{fig_EDL}, indicating a nontrivial topological index. This suggests a weak interaction cannot lead to a transition to the trivial phase, which is in agreement with Ref.~\onlinecite{HOTIC2}.}

\subsection{Stability against boundary deformation }

The gapless hinge modes emerge as features indicating the higher-order topology. This nomenclature seems to suggest the requirement of a sharp hinge at the boundary, and it raises a natural question as to whether these hinge modes still exist if the boundary does not have a sharp hinge. In addition, the bulk-boundary correspondence (BBC) for symmetry-protected higher-order topological phases usually requires the set of boundaries to preserve the same symmetry~\cite{Bernevig_HOTI,Trifunovic2021}, e.g., the four side surfaces in Fig.\ref{fig_hinge}(a) for the model in Eq.\eqref{HOTImodel} need to be related to each other by fourfold rotational symmetry. This is because the hinge is the direct boundary of 2D surfaces rather than the 3D bulk. If 2D surfaces are allowed to break the symmetry, one can annihilate the hinge states without closing the 3D bulk gap. In reality the crystalline symmetry can be easily broken by the boundary truncation, and it is natural to ask whether these hinge modes still exist. {In this section we aim to answer whether the existence of hinge states requires the boundaries as a whole to preserve the symmetry, and whether it requires the boundary to have a sharp physical hinge. We show that neither of these conditions are necessary for the hinge states to appear.}

\begin{figure}
\includegraphics[width=3.3 in]{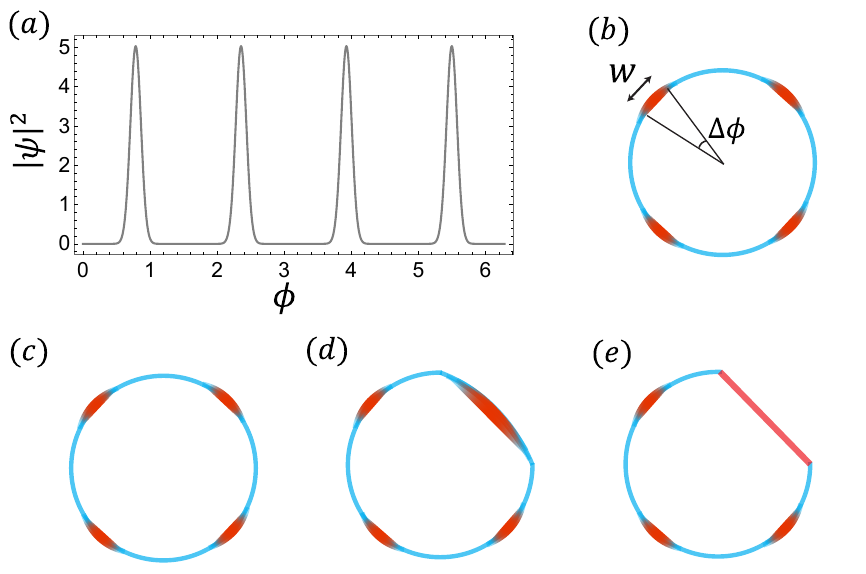}
\caption{ (a): There are four eigenstates of the effective surface Hamiltonian at zero energy located at $\phi=\pm\frac{\pi}{4},\pm\frac{3\pi}{4}$. Here $|\psi|^2=|a(\phi)|^2+|b(\phi)|^2,\ m_s=1,\ R=20$. (b): Top view of the cylindrical boundary and the four effective hinge modes. The angular width is $\Delta\phi$ and the linear width is $w$. (c)-(e): Evolution of the effective hinge mode under deformation of the boundary. When the boundary is flattened, the linear width of the hinge mode increases and it becomes a surface mode when the boundary becomes flat.
   }
\label{fig_cyhinge}
\end{figure}

Consider a sample that microscopically realizes a higher-order topological insulator, e.g., the model in Eq.\eqref{HOTImodel} and possibly with interactions, but has the macroscopic shape of a cylinder with axis along $z$ direction, instead of the cubic shape in Fig.\ref{fig_hinge}(a). If there is no $\Delta_2$ term, the system is a first-order topological insulator and the effective Hamiltonian on the side surface of the cylinder is a gapless Dirac cone described by 
\be
\tilde H_{s}=k_\phi \sigma_1+k_z\sigma_2.
\ee
Here $k_\phi$ is the momentum component along the circumferential direction. The $\Delta_2$ term breaks $T$ and $C_4$ separately and generates a surface mass term with angle dependence $\cos 2\phi$ such that it flips sign under fourfold rotation. The effective Hamiltonian for the surface becomes
\be
H_s=-i\frac{1}{R}\frac{\partial}{\partial\phi}\sigma_1+k_z\sigma_2+m_{s}\cos 2\phi \sigma_3.
\ee
Here $R$ is the radius of the cylinder. {In the thermodynamic limit $R\rightarrow\infty$, the surface mass approaches the value of a flat surface and does not depend on $R$.} The system is still periodic along $z$ direction and the $z$-dependence of eigenstates are of the usual Bloch form $e^{ik_z z}$. To find boundary modes with zero energy, we focus on $k_z=0$ and solve the following differential equation to obtain the eigenstates of the surface Hamiltonian:
\bea
\begin{pmatrix}
m_s \cos 2\phi & -i\frac{1}{R}\frac{\partial}{\partial\phi} \\
-i\frac{1}{R}\frac{\partial}{\partial\phi} & -m_s \cos 2\phi
\end{pmatrix}
\begin{pmatrix}
a(\phi) \\
b(\phi)
\end{pmatrix}=E\begin{pmatrix}
a(\phi) \\
b(\phi)
\end{pmatrix}.
\label{Eqsurf}
\eea
This equation can be solved by expanding $a(\phi)$ and $b(\phi)$ in terms of $e^{\pm in\phi}$. The solution shows that for large $R$ there are four eigenstates with zero energy located at $\phi=\pm\frac{\pi}{4},\pm\frac{3\pi}{4}$, as in Fig.\ref{fig_cyhinge}(a),(b). These eigenstates are close to Gaussian functions of $\phi$ and are spatially separated. The small angular width $\Delta\phi$ and the extended nature along $z$ direction make these eigenstates effectively "hinge states" although the cylindrical boundary has no hinge.

\begin{figure}
\includegraphics[width=3.4 in]{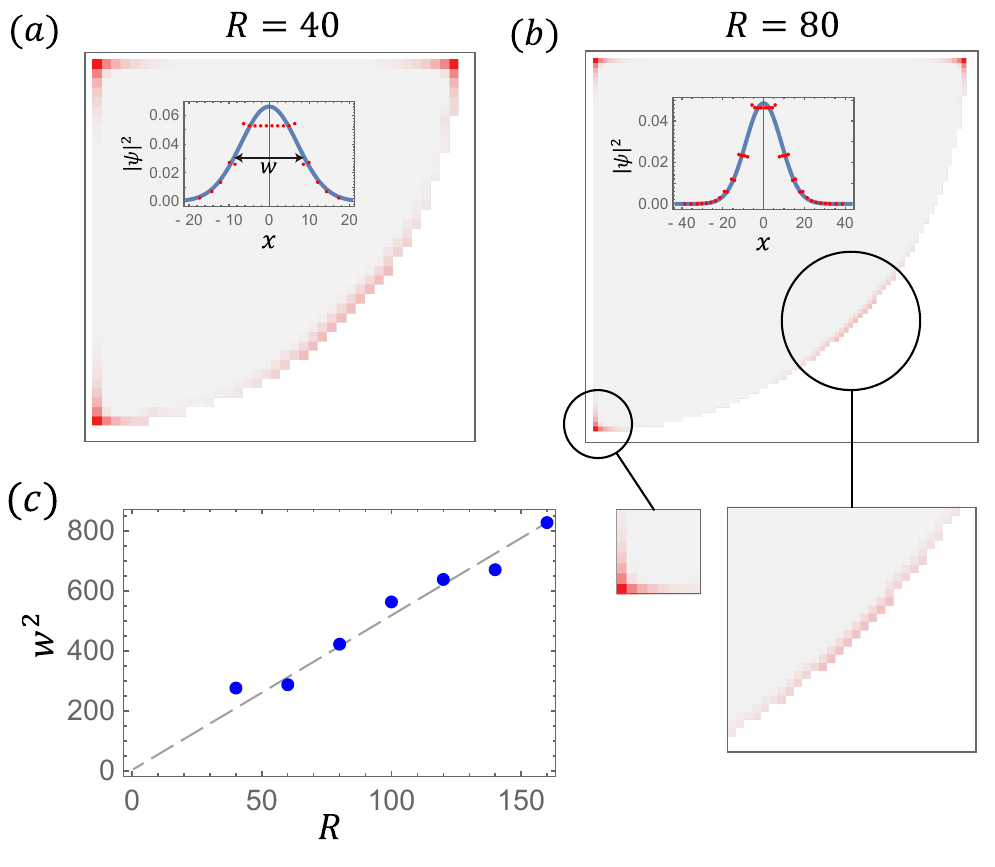}
\caption{ {Wave functions of the model in Eq.\eqref{HOTImodel} solved on a lattice of one-quarter-cylinder geometry with radius $R=40$ in (a) and $R=80$ in (b). The four eigenstates with zero energy at $k_z=0$ are plotted. The parameters are chosen as $t_x=t_y=t_z=1,M=-2,\Delta_x=\Delta_y=\Delta_z=1,\lambda_1=\lambda_2=0,\Delta_2=0.3$. The inset shows the Gaussian fit $|\psi|^2=A\exp(-\frac{x^2}{2\sigma^2})$ of the wave function of the zero mode at the cylindrical surface, where $x$ is the distance to the center of the mode. The width $w$ is obtained from the fit by $w=2.35\sigma$. $w^2$ as a function of $R$ for different system sizes is shown in (c). The linear relation between $w^2$ and $R$ indicates $w\sim \sqrt{R}$ and $\Delta\phi=w/R\sim 1/\sqrt{R}$. Therefore, as $R$ increases the angular width $\Delta\phi$ decreases and the state becomes an effective hinge state although there is no sharp hinge at the boundary.}
   }
\label{fig_hingenum}
\end{figure}

The emergence of these effective hinge states can be understood by linearly expanding the surface Hamiltonian near $\phi=\pm\frac{\pi}{4},\pm\frac{3\pi}{4}$ where the surface mass changes sign. Let $x$ denote the spatial coordinate along $\phi$ direction, Eq.\eqref{Eqsurf} reduces to
\bea
\begin{pmatrix}
\pm\frac{2m_s x}{R} & -i\frac{\partial}{\partial x} \\
-i\frac{\partial}{\partial x} & \mp\frac{2m_s x}{R}
\end{pmatrix}
\begin{pmatrix}
a(x) \\
b(x)
\end{pmatrix}=0,
\label{Eqsurfl}
\eea
{The linearized equation Eq.\eqref{Eqsurfl} can be solved at $\phi=\pm\frac{\pi}{4},\pm\frac{3\pi}{4}$ independently. It has a simple solution $(a,b)\sim(1,\pm i)e^{-\frac{x^2}{R/m_s}}$. If local perturbations are added to one hinge, it cannot affect the other hinges that are spatially separated by a macroscopic distance, hence the other hinges still follow Eq.\eqref{Eqsurfl} with a solution of hinge mode. This indicates the boundaries does not need to preserve the $C_4$ symmetry for the hinge modes to appear.} The factor $e^{-\frac{x^2}{R/m_s}}$ shows these eigenstates are Gaussian functions with linear width $w\sim\sqrt{\frac{R}{m_s}}$ and angular width $\Delta\phi=w/R\sim\frac{1}{\sqrt{Rm_s}}$. The fact that $\Delta\phi\sim\frac{1}{\sqrt{R}}$ indicates $\Delta\phi\rightarrow 0$ in the thermodynamic limit $R\rightarrow\infty$, hence as the system size increases the hinge states will be localized in a smaller angular range even if the boundary does not have a sharp hinge. The behavior of linear width $w\sim \sqrt{R}$ at large $R$ is also consistent with the evolution of hinge states under the flattening of cylindrical surface. Consider the flattening process in Fig.\ref{fig_cyhinge}(c)-(e). From (c) to (e) the surface radius increases to infinity under the flattening process. According to $w\sim \sqrt{R}$, the linear width of the hinge state will increase and occupy the whole surface when the surface becomes flat. This is consistent with the fact that the surface mass term $m_s\cos 2\phi$ can gap out the surfaces perpendicular to $x$ or $y$ directions but not the surfaces perpendicular to $\hat x+\hat y$, {where $x$ and $y$ directions are along crystalline axes corresponding to $\phi=0$ and $\phi=\pi/2$ respectively.}

To verify these predictions obtained from the effective surface Hamiltonian, we diagonalize the tight-binding model in Eq.\eqref{HOTImodel} on a finite-size system with the shape of one-quarter cylinder in Fig.\ref{fig_hingenum}. The $z$ direction is taken to be periodic. We find that at $k_z=0$, there are four eigenstates with zero energy, with three of them localized at the hinges, and one localized at cylindrical part of the surface. {We use Gaussian fit $|\psi|^2=A\exp(-\frac{x^2}{2\sigma^2})$ to obtain the width of the state at the cylindrical surface, as shown in the inset of Fig.\ref{fig_hingenum}. Here $x$ is the distance to the center of the zero mode. Define width $w$ by the location where the Gaussian curve reduces to half its peak value, then $w=2.35\sigma$. The fitting shows the wave function profile approaches a Gaussian function when $R$ increases, while for small $R=40$ there is deviation from Gaussian due to finite size effect. For $R=40$ and $80$ we obtain $w=16.6$ and $20.6$ respectively. $w^2$ as a function of $R$ in various system sizes is shown in Fig.\ref{fig_hingenum}(c). The linear relation between $w^2$ and $R$ indicates $w\sim \sqrt{R}$ and $\Delta\phi\sim 1/\sqrt{R}$ at large system size.} Therefore, as the system size increases, the angular width $\Delta\phi$ of the state at the cylindrical surface decreases, approaching an effective hinge state. Therefore, the emergence of hinge states as features of higher-order topology does not require the boundaries to preserve the protecting symmetry nor does it require the existence of a physical hinge. {Although the computation in Fig.\ref{fig_hingenum} is performed for a non-interacting system, the hinge modes are expected to persist under weak interaction due to the presence of a finite surface gap. However, a direct numerical verification of the existence of hinge states in an interacting system with large system size is beyond the scope of this work.}

We emphasize that the above conclusions do not imply these hinge states will emerge under arbitrary boundary conditions that break the protecting symmetry. For example, due to the absence of $C_4T$ symmetry on the boundary, in principle the hinge states in Fig.\ref{fig_hinge}(a) can be removed by superimposing a layer of Chern insulator to the left and right side surfaces respectively. However, this process has to be a large perturbation that closes the surface gap. {Our results in Fig.\ref{fig_cyhinge} and \ref{fig_hingenum} show that if the boundary is obtained by truncating the crystal rather than decorating with another material, then the presence of hinge states is a local property of the boundary, independent of whether the boundary satisfies the global symmetry. In regards of the realization of these hinge states in experiments, this implies the boundary termination of the sample does not need to strictly obey the symmetry and does not require a sharp hinge for these hinge states to be observed.}



\section{Conclusion}

We show that the eigenstates of inverse Green's function at zero frequency are useful tools in characterizing higher-order topological phases with electronic interaction. In particular, it enables us to compute the topological index of interacting $C_4T$-symmetric second-order topological insulator in a gauge-independent way, and with additional $S_4$ symmetry the topological index the interacting system can be determined by $S_4$ eigenvalues directly, similar to the Fu-Kane formula. This  Green's function-based approach can also be applied to compute the topological index for higher-order topological superconductors. We demonstrate that the hinge states as features of higher-order topology are robust to interaction and deformation of boundaries. If the sample of higher-order topological insulator is sufficiently large and has natural open boundary condition such that its boundaries are obtained by truncating the crystal, the hinge states survive even in the absence of a physical hinge at the boundary. We also propose crystal structures with $C_4T$-preserving magnetic order as possible platforms to realize this higher-order topological phase with electron interactions.

\section{Acknowledgement}

This work is supported by the Natural Sciences and Engineering Research Council of Canada (NSERC) and the Center for Quantum Materials at the University of Toronto. H.Y.K acknowledges the support by the Canadian Institute for Advanced Research (CIFAR) and the Canada Research Chairs Program.

\appendix

\setcounter{equation}{0}
\setcounter{figure}{0}
\setcounter{table}{0}
\makeatletter
\renewcommand{\theequation}{S\arabic{equation}}
\renewcommand{\thefigure}{S\arabic{figure}}

\section{Details in the derivation of topological indices}

\subsection{Proof of anti-symmetry}

The Pfaffian formula Eq.\eqref{Pfmain} requires the matrix $M$ to be anti-symmetric. We show that $M(\mathbf k)$ is anti-symmetric for every $C_4$-invariant $\mathbf k$ point such that $C_4T\mathbf k=C_4^{-1}T\mathbf k=-\mathbf k$. Let $\Theta=\frac{C_4T+C_4^{-1}T}{\sqrt{2}}$, we can show that $\Theta^2=T^2=-1$. This is due to $(C_4T)^4=(C_2)^2=-1$, then $C_2+C_2^{-1}=0$ and $\Theta^2=-(C_4+C_4^{-1})^2/2=-(C_2+C_2^{-1}+2)/2=-1$. Therefore $\Theta$ is an anti-unitary operator similar to the time-reversal operator for spin 1/2 systems which gives rise to Kramers degeneracy. Then for every $C_4$-invariant momentum $\mathbf k$ we have
\bea
M_{mn}(\mathbf{k})&=&\bra{g_m(\mathbf{k})} \Theta\ket{g_n(\mathbf{k})} \nonumber\\
&=&\bra{\Theta^2 g_n(\mathbf{k})} \Theta\ket{g_m(\mathbf{k})}\nonumber\\
&=&-\bra{g_n(\mathbf{k})} \Theta\ket{g_m(\mathbf{k})}\nonumber\\
&=&-M_{nm}(\mathbf{k}).
\eea
This shows $M(\mathbf k)$ is anti-symmetric, hence its Pfaffian is well-defined.

\subsection{Proof of Eq.\eqref{P3ginv}}

We present the derivation that leads to Eq.\eqref{P3ginv} in the main text. This part follows Ref. \onlinecite{LiTIC2T}. First we show that the line quantity $\mathcal R$ in Eq.\eqref{gwilson} is gauge-invariant, i.e., invariant under gauge transformation $|g_n(\mathbf k)\rangle\rightarrow\sum_m |g_m(\mathbf k)\rangle U_{mn}(\mathbf k)$ where $U$ is a unitary matrix and the summation is over the eigenstates of inverse Green's function with positive eigenvalues. First consider the limit in which $\mathbf k_1$ and $\mathbf k_2$ are close to each other. Then $W(\mathbf k_1,\mathbf k_2)=\langle g_m(\mathbf k_1)\ket{g_n(\mathbf k_2)}$. Under the gauge transformation, $M_{mn}(\mathbf k)=\langle g_m(\mathbf k)| \Theta|g_n(\mathbf k) \rangle\rightarrow (U(\mathbf k)^\dagger M(\mathbf k) U(\mathbf k)^*)_{mn}$, $W_{mn}(\mathbf k_1,\mathbf k_2)=\langle g_m(\mathbf k_1)|g_n(\mathbf k_2)\rangle \rightarrow (U(\mathbf k_1)^\dagger W(\mathbf k_1,\mathbf k_2) U(\mathbf k_2))_{mn}$. Therefore
\bea
\operatorname{Pf}[M(\mathbf k)]&\rightarrow& \operatorname{Pf}[M(\mathbf k)] \det[U(\mathbf k)]^* ,  \nonumber\\
\det[W(\mathbf k_1,\mathbf k_2)]&\rightarrow&\det[W(\mathbf k_1,\mathbf k_2)]\det[U(\mathbf k_1)]^* \det[U(\mathbf k_2)], \nonumber\\
\mathcal R(\mathbf k_1,\mathbf k_2)&=&\frac{\operatorname{Pf}[M(\mathbf k_2)]}{\operatorname{Pf}[M(\mathbf k_1)]}\det[W(\mathbf k_1,\mathbf k_2)]\rightarrow \mathcal R(\mathbf k_1,\mathbf k_2). \nonumber
\eea
This shows that $\mathcal R(\mathbf k_1,\mathbf k_2)$ is gauge-invariant when $\mathbf k_1$ is close to $\mathbf k_2$. For a general pair of separated momentum points $\mathbf k_a$ and $\mathbf k_b$, we can divide the path connecting $\mathbf k_a$, $\mathbf k_b$ by small segments $(\mathbf k_i,\mathbf k_{i+1})$. Then $\mathcal R(\mathbf k_a,\mathbf k_b)=\prod_i \mathcal R(\mathbf k_i,\mathbf k_{i+1})$. For each small segment $R(\mathbf k_i,\mathbf k_{i+1})$ is gauge-invariant, therefore $\mathcal R(\mathbf k_a,\mathbf k_b)$ is gauge-invariant as well. This gauge-invariance allows us to compute $\mathcal R$ without a smooth gauge.

Next we show that $\mathcal R$ can be related to the topological index $P_3$. Select a gauge on the straight line $\overline{ZZ'}$ such that $\ket{g_n(\mathbf k)}$ is smooth and periodic. For each $\mathbf k=(k_x,k_y,k_z)$ in the region $\tau_{1/2}$ in Fig.\ref{fig_hinge}(b), let $\mathbf k_p=(0,0,k_z)$ and then a parallel transport gauge~\cite{smoothgauge} that is smooth in $\tau_{1/2}$ can be defined by
\be
\ket{g_m(\mathbf k)}=\prod_{\mathbf k_i\in \overline{\mathbf k \mathbf k_p}}^{\mathbf k\leftarrow \mathbf k_p} P_{\mathbf k_i} \ket{g_m(\mathbf k_p)}.
\label{pgauge}
\ee
In this gauge for each $\mathbf k\in \overline{ZZ'}$, $\det[W(\mathbf k,\mathbf k+\mathbf G_0)]$ becomes unity, and we get
\be
\mathcal R(\mathbf k,\mathbf k+\mathbf G_0)=\frac{\operatorname{Pf}[M(\mathbf k+\mathbf G_0)]}{\operatorname{Pf}[M(\mathbf k)]}.
\ee
Note that $\mathbf k+\mathbf G_0\in \overline{AA'}$ if $\mathbf k\in \overline{ZZ'}$. Therefore, Eq.\eqref{P3ginv} becomes the winding of the phase of $\operatorname{Pf}[M(\mathbf k)]$ along $\overline{ZZ'}$ and $\overline{AA'}$, which is equivalent to $\partial\tau_{1/2}$ since contribution along $\overline{ZA}$ and $\overline{Z'A'}$ cancel by periodicity. Therefore, Eq.\eqref{P3ginv} represents the winding of Pfaffian along $\partial\tau_{1/2}$, which is equivalent to Eq.\eqref{Pfmain}. This proves Eq.\eqref{P3ginv} in the parallel transport gauge. Because $\mathcal R$ is gauge-invariant, Eq.\eqref{P3ginv} should be true in any gauge. This finishes the proof.

The advantage of Eq.\eqref{P3ginv} is that because $\mathcal R$ is gauge-invariant, the construction of the smooth parallel transport gauge in Eq.\eqref{pgauge} is not needed in actual computation. In practice one can evaluate Eq.\eqref{P3ginv} in any gauge obtained by diagonalizing the inverse Green's function.


%

\end{document}